\documentclass[aps,prx,amsfonts,floatfix,superscriptaddress,longbibliography,twocolumn,footinbib]{revtex4-2}

\usepackage[utf8]{inputenc}
\usepackage{amsmath}    
\usepackage{amssymb}
\usepackage{mathbbol}  
\usepackage{verbatim}   
\usepackage{subfigure} 
\usepackage{scalerel}
\usepackage{soul}
\usepackage{siunitx}
\usepackage{physics}
\usepackage{enumerate} 
\usepackage{empheq}

\usepackage{graphicx}% Include figure files
\usepackage{dcolumn}% Align table columns on decimal point
\usepackage{bm}% bold math
\usepackage[colorlinks = True, linkcolor = red, citecolor = red]{hyperref}
\usepackage{graphicx}
\usepackage[normalem]{ulem}
\usepackage{xcolor}
\usepackage{enumitem}
\usepackage{orcidlink}

\usepackage{tikz}
\usepackage{pgfplots}
\usetikzlibrary{shapes, arrows, positioning, calc}
\pgfplotsset{compat=1.18}

\definecolor{panelblue}{HTML}{002D62}
\definecolor{panelgreen}{HTML}{004B23}
\definecolor{panelpurple}{HTML}{4A0E4E}
\definecolor{panelorange}{HTML}{D9541E}

\definecolor{curveblue}{HTML}{1A73E8}
\definecolor{curvered}{HTML}{D93025}
\definecolor{curvegreen}{HTML}{1E8E3E}

\tikzset{
    axis/.style={thick, ->, >=stealth, black},
    label/.style={font=\fontfamily{phv}\selectfont\small},
    sublabel/.style={font=\fontfamily{phv}\selectfont\footnotesize\itshape}
}
\begin{document}

\title{Fragmented ETH and Ensemble Inequivalence}

\author{C. L. Sriram\,\orcidlink{0009-0001-3706-6498}} 
%\email{c.l.sriram@uconn.edu}
\affiliation{Department of Physics, University of Connecticut, Storrs, Connecticut, USA}

\author{Soumya Kanti Pal\,\orcidlink{0009-0008-7226-356X}} 
%\email{soumya.pal@tifr.res.in} 
\affiliation{Department of Theoretical Physics, Tata Institute of Fundamental Research, Homi Bhabha Road, Mumbai 400005, India}

\author{Lea F. Santos\,\orcidlink{0000-0001-9400-2709}} 
%\email{lea.santos@uconn.edu} 
\affiliation{Department of Physics, University of Connecticut, Storrs, Connecticut, USA}

\begin{abstract}
We investigate thermalization in finite quantum systems with strong long-range interactions. Nearly conserved quantities inherited from the fully connected limit organize the Hilbert space into weakly coupled sectors and split the many-body spectrum into energy bands.
Despite the resulting breakdown of global ergodicity, quantum chaos develops within individual energy bands, enabling the definition of microcanonical ensembles within the bands. This supports a band-resolved formulation of thermalization, which we term fragmented eigenstate thermalization hypothesis (fETH). Unlike conventional ETH, finite-size scaling in fETH obeys a symmetry-imposed selection rule that restricts which system sizes can be compared.
This band-resolved description has consequences for equilibrium statistical mechanics. While microcanonical ensembles remain confined to a single band, canonical ensembles mix different bands. This mismatch explains ensemble inequivalence without invoking equilibrium phase transitions. Our results apply broadly to Hamiltonians near a fully permutation-symmetric limit.
\end{abstract}

\maketitle

\section{Introduction}
\label{sec:Introduction}

A fundamental question in statistical physics is how thermodynamics emerges from microscopic principles. In particular, can a pure quantum state evolving unitarily exhibit equilibrium behavior consistent with thermodynamics? This question was addressed by von Neumann~\cite{vonNeuman1929, vonNeumann1929b}  and later explored in the context of quantum chaos, where chaotic eigenstates display random-matrix-like properties consistent with thermal equilibrium~\cite{Jensen1985,Deutsch1991,Srednicki1994,Srednicki1996,ZelevinskyRep1996,Flambaum1997,Santos2010PRE,Borgonovi2016,Alessio2016}. The modern understanding is largely based on the framework of the eigenstate thermalization hypothesis (ETH) \cite{Srednicki1994}, which concerns the matrix elements of few-body observables in the energy eigenbasis. ETH asserts that when the diagonal matrix elements of the observables are smooth functions of energy and the off-diagonal elements are exponentially suppressed in system size, the observable expectation values in individual eigenstates coincide with thermodynamic predictions~\cite{Alessio2016}.

In this work, we investigate how thermalization is modified in finite quantum systems with strong long-range interactions. Systems with long-range interactions~\cite{Defenu2023,Defenu2024}, where couplings $J_{ij}$ between sites $i$ and $j$ decay as a power law of their distance, $J_{ij} \propto 1/|i-j|^\alpha$~\cite{Sadugol2026}, are now routinely realized in experiments with trapped ions~\cite{Lanyon2011,Jurcevic2014,Neyenhuise2017,Britton2012,Richerme2014}, Rydberg atoms~\cite{Saffman2010,Scholl2021}, and cavity quantum electrodynamics setups~\cite{Li2023,Luo2025}. We focus on the super long-range regime, $\alpha<d$, where $d$ is the dimension of the system.

In this regime, the nearly conserved quantities inherited from the fully connected limit ($\alpha=0$) partition the Hilbert space into approximate symmetry sectors and split the many-body spectrum into energy bands. This leads to the breakdown of global ergodicity and anomalously slow equilibration~\cite{Sriram2026_2}, which has motivated suggestions that ETH fails in strongly long-range interacting systems~\cite{Russomanno2021,Sugimoto2022}.

The fragmentation of the Hilbert space into weakly coupled sectors raises two fundamental questions that we address here. Can ETH remain valid when global ergodicity is lost? How does the emergence of energy bands affect the statistical description of thermal equilibrium and, in particular, the equivalence between microcanonical and canonical ensembles?

We demonstrate that the breakdown of global ergodicity and the failure of ETH when applied to the full spectrum do not imply the breakdown of thermalization. As soon as the system departs from the fully connected limit, however weakly, the degeneracies within each energy band are lifted and the intraband eigenvalues become correlated, signaling the onset of quantum chaos. This intraband chaos enables thermalization to occur separately within each band, leading to a band-resolved formulation of ETH that we term fragmented ETH (fETH)~\cite{Pal2025}.

At finite interaction range this fragmentation is only approximate, since weak hybridization between neighboring bands induces leakage that grows with system size. We develop a scaling theory for this leakage, and establish the parameter regime in which thermalization is properly governed by fETH. Within this regime, we show that the eigenstate-to-eigenstate fluctuations of local observables decrease with increasing system size, supporting the validity of fETH.

Testing fETH requires a finite-size scaling analysis that differs from that used in conventional ETH, where system sizes are typically taken consecutively or in steps of two. Because the spectral band structure originates from the permutation symmetry of the fully connected parent Hamiltonian, bands at different system sizes can be compared only when they are symmetry compatible. We rigorously prove that this permutation symmetry, together with spin-inversion symmetry, imposes a selection rule on comparable system sizes, restricting the scaling analysis to sizes $L,L+4,L+8,\ldots$.

The band-resolved nature of fETH also leads to a new perspective on ensemble inequivalence.  In short-range systems, the equivalence between microcanonical and canonical ensembles relies on additivity, whereby interactions between subsystems become negligible in the thermodynamic limit. This property breaks down in super-long-range interacting systems, where interactions remain significant across the entire system~\cite{RuffoBook, Dauxois2002, Campa2009, Campa2014}. Ensemble inequivalence is commonly discussed in connection with equilibrium phase transitions~\cite{Barre2001, Defenu2024_1,ArrufatVicente2025}. Here, instead, we identify a microscopic mechanism rooted in the band structure of the many-body spectrum. While the microcanonical ensemble is confined to a single energy band, the canonical ensemble samples states from multiple bands. Since the two ensembles explore different regions of Hilbert space, they generally yield different equilibrium predictions. This inequivalence is reflected in the caloric curves. The canonical relation between temperature and energy is single valued, whereas the band-resolved microcanonical relation becomes multi-valued across the fragmented spectrum.

The highly degenerate band structure underlying the results presented here is a generic consequence of full permutation symmetry and therefore arises in a broad class of Hamiltonians. As a representative example, we illustrate our findings using the experimentally relevant spin-$1/2$ transverse-field Ising chain with strong long-range interactions.

The paper is organized as follows. Section~\ref{sec:Model} introduces the model and discusses the emergence of spectral fragmentation and quantum chaos within individual energy bands. Section~\ref{sec:Leakage} quantifies interband hybridization and establishes the regime in which a band-resolved statistical description remains valid. Section~\ref{sec:MC} develops the fETH framework and the symmetry-imposed finite-size scaling required to test it. Section~\ref{sec:Ensemble_Inequivalence} examines the consequences of fETH for the equivalence between microcanonical and canonical ensembles. We summarize our conclusions in Sec.~\ref{sec:Conclusion}.

%%%%%%%%%%%%%%%%%%%%%%%%%%%%%%%%%%%

%%%%%%%%%%%%%% MODEL %%%%%%%%%%%%%%%
\section{Energy Bands and Quantum Chaos}
\label{sec:Model}

We consider the one-dimensional (1D) spin-$1/2$ transverse field Ising model with two-body interactions that decay with distance as a power law. The system is described by the Hamiltonian
\begin{align}
\hat{H}^{(\alpha)} = \sum_{i<j} \frac{J}{\mathcal{N}_\alpha} \frac{\hat{\sigma}_i^x \hat{\sigma}_j^x}{|i-j|^\alpha}
+ h \sum_{i=1}^{L} \hat{\sigma}_i^z,
\label{eq:XXtZH}
\end{align}
where $\hat{\sigma}_i^{x,y,z}$ are Pauli matrices at site $i$, $L$ is the system size, $J$ sets the interaction scale, and $h$ is the transverse magnetic field strength. Throughout this work, we fix $J=1$ and $h=1$. We investigate the case of super-long-range interactions characterized by $\alpha<1$. To ensure the extensivity of the Hamiltonian, we include the Kac's factor $\mathcal{N}_\alpha \sim L^{1-\alpha}$.  We denote the eigenvalues and eigenstates of $\hat{H}^{(\alpha)}$ for $\alpha \neq 0$ as 
\begin{equation}
    \hat{H}^{(\alpha)} |n \rangle = E_n |n \rangle .
    \label{Eq:EigenstatesH}
\end{equation}

For any value of $\alpha$, the Hamiltonian in Eq.~\eqref{eq:XXtZH} is invariant under both parity and spin inversion. Parity corresponds to the spatial reflection of the chain, $i \to L+1-i$, that is, $\hat{P}\,\hat{\sigma}_i^{x,y,z}\,\hat{P}^{-1} = \hat{\sigma}_{L+1-i}^{x,y,z} $. Spin inversion corresponds to a global flip of all spins and is generated by the operator $\hat{R}^z = \prod_i \hat{\sigma}_i^z$. These symmetry operators, $\hat{P}$ and $\hat{R}^z$, commute with the Hamiltonian, so the Hilbert space decomposes into sectors labeled by their eigenvalues $p = \pm 1$ and $r = \pm 1$. This implies that the dimension of each sector is approximately ${\cal D}_{p,r} \sim 2^L/4 $.

The model exhibits two integrable limits that are experimentally accessible: 
\begin{enumerate}
 \item  For $\alpha \to \infty$, Eq.~\eqref{eq:XXtZH} reduces to the nearest-neighbor transverse-field Ising chain. 
 
  \item At $\alpha = 0$, it reaches the fully connected regime, where any pair of spins interacts with equal strength. Such collective spin systems are realized in cavity quantum electrodynamics setups~\cite{Li2023,Luo2025}. 
  In this all-to-all limit, the Hamiltonian coincides with that of the Lipkin–Meshkov–Glick (LMG) model and can be written as
\begin{align}
\hat{H}^{(0)}_{\text{LMG}} = \frac{2J}{N} \hat{S}_x^2 + 2h \hat{S}_z - 2J,
\label{Eq:LMG}
\end{align}
where $\hat{S}_{x,y,z} = \frac{1}{2} \sum_i \hat{\sigma}_i^{x,y,z}$ are collective spin operators. The LMG Hamiltonian possesses a global $\mathrm{SU}(2)$ symmetry,
\begin{align}
[\hat{S}^2, \hat{H}^{(0)}_{\text{LMG}}] = 0,
\end{align}
where $\hat{S}^2 = \hat{S}_x^2 + \hat{S}_y^2 + \hat{S}_z^2$ is the total spin angular momentum. We denote the eigenvalues and eigenstates of $\hat{H}_{\text{LMG}}^{(0)}$  as 
\begin{equation}
    \hat{H}^{(0)}_{\text{LMG}} |n^{(0)} \rangle = E_n^{(0)} |n^{(0)} \rangle .
    \label{Eq:EigenstatesH0}
\end{equation}

The spectrum of the LMG model is organized into families of energy bands labeled by the total spin quantum number $s$. For a system of even size $L$, the total spin takes values $s=0,1,\ldots,L/2$, where $s = L/2$ corresponds to the fully symmetric sector. Due to the transverse field, each $s$-sector further splits into $(2s+1)$ distinct energy bands. The total number of bands is therefore
\begin{equation}
\sum_{s=0}^{L/2} (2s+1) = \left( \frac{L}{2} + 1 \right)^2.
\end{equation}

The number of states in each band depends on $s$ and $L$. It equals the dimension of the
multiplicity space and is given by
\begin{align}
g(s,L) = \frac{2s + 1}{L + 1} \binom{L + 1}{\frac{L}{2} - s},
\label{Eq:gsL}
\end{align}
which follows from the combinatorics of angular momentum addition. These bands are highly degenerate, except for the fully symmetric sector,
$s=L/2$. In this case,
$g(L/2,L)=1$, so each of the $(L+1)$
energy bands contains a single eigenstate. 

\end{enumerate}

\begin{figure}[h]
    \begin{center}
        \includegraphics[width = 0.7\columnwidth]{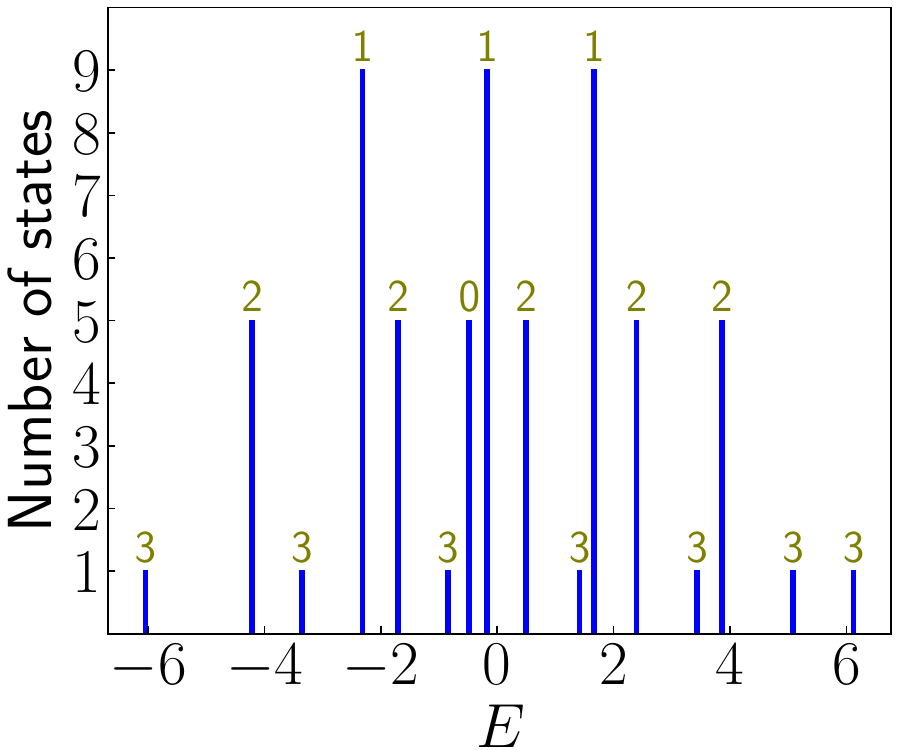}
        \caption{Density of states for the whole spectrum of the LMG Hamiltonian $\hat{H}^{(0)}_{\text{LMG}}$ in Eq.~(\ref{Eq:LMG}) for $L=6$ sites. The value of $s$ for each band is indicated in red. 
        }
        \label{fig:FIG1}
    \end{center}
\end{figure}

The band structure of the fully connected limit, $\alpha=0$, is shown in Fig.~\ref{fig:FIG1}(a) for $L=6$. There are 16 energy bands, with populations ranging from just one level for each of the $7$ bands with $s=L/2$ to 9 degenerate levels for each of the three bands with $s=1$. 

In Figs.~\ref{fig:FIG2}(a)-(b) we analyze the DOS for the model with super-long-range interactions at $\alpha =10^{-4}$ and $L=16$. As shown in Fig.~\ref{fig:FIG2}(a), $\hat{H}^{(\alpha)}$ retains the band structure of the LMG Hamiltonian for $\alpha<1$. For any infinitesimally small value of $\alpha$, the degeneracies of the LMG limit are lifted and the DOS within individual bands acquires a Gaussian profile, as shown in  Fig.~\ref{fig:FIG2}(b).  This shape is typical of many-body systems with two-body interactions.  

The band structure of the spectrum plays a central role in our analysis. We focus on the regime $0 < \alpha \ll 1$, where the Hamiltonian can be viewed as a perturbation of the fully connected limit. 
%%%%%%%%%%%%%%%%%%%%%%%%%%%%
\iffalse
\begin{align}
\hat{H}^{(\alpha)} = \hat{H}^{(0)} + \epsilon \hat{V}_\alpha .
\label{Eq:PertH_H0_V}
\end{align}
In the equation above, $\hat{V}_\alpha$ introduces nonlocal corrections that break permutation symmetry and $\epsilon$ characterizes the strength of the perturbation. 
\fi
%%%%%%%%%%%%%%%%%%%%%%%%%%%%

\subsection{Quantum Chaos}
\label{Sec:QChaos}

Despite the band structure of the DOS, the spectral statistics of the system reveals signatures of chaos within individual energy bands~\cite{Russomanno2021}. The onset of quantum chaos is thus fragmented across the spectrum~\cite{Pal2025}.

Quantum chaos is characterized by correlated eigenvalues as described by random matrix theory~\cite{MehtaBook,Guhr1998}. Short-range spectral correlations, in particular, can be quantified using the distribution of ratios of adjacent level spacings~\cite{Atas2013}
\[
r_n = \frac{\min({\cal S}_n,{\cal S}_{n-1})}{\max({\cal S}_n,{\cal S}_{n-1})},
\qquad
{\cal S}_n = E_{n+1} - E_n .
\]
Level repulsion results in the Wigner–Dyson distribution, whereas integrable systems are characterized by uncorrelated spectra with Poissonian level statistics or by picket-fence spectra.  

\begin{figure}
    \begin{center}
        \includegraphics[width = 1.0\columnwidth]{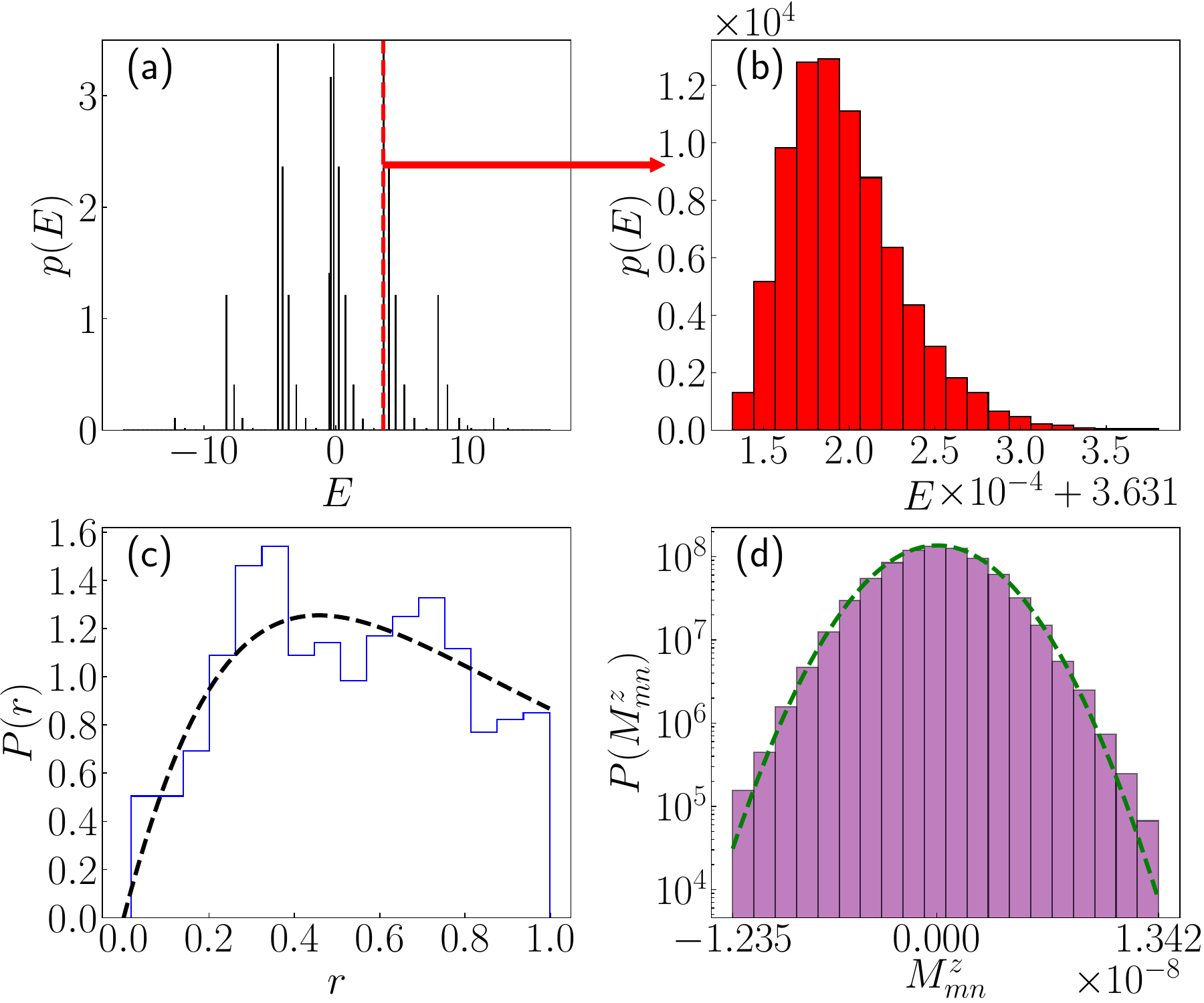}
        \caption{(a) Density of states (DOS) of the full spectrum of $\hat{H}^{(\alpha)}$ in Eq.~(\ref{eq:XXtZH}), showing that the band structure of the fully connected LMG model persists when $0<\alpha<1$. The vertical dashed red line indicates the band which is magnified in (b) and is considered also in the analysis of quantum chaos in (c) and (d). (b) DOS within the selected energy band, exhibiting a skewed Gaussian shape. (c) Distribution of the ratio of consecutive level spacings. (d)  Distribution of the off-diagonal elements of the total $z$-magnetization $M^z$; 264 eigenstates are selected in the middle of the energy band. All panels: $L=16$, $\alpha=10^{-4}$.
        }
        \label{fig:FIG2}
    \end{center}
\end{figure}

In Fig.~\ref{fig:FIG2}(c), we show the distribution of the ratio of consecutive level spacings, $r_n$, for one of the most populated bands of the spectrum of $\hat{H}^{(\alpha)}$ with $\alpha =10^{-4}$, finding excellent agreement with the Wigner-Dyson distribution. This indicates that any infinitesimal value of $\alpha$ lifts the degeneracies of the collective LMG model and induces correlations between eigenvalues, even for finite system sizes. This behavior is in contrast with that of systems featuring short-range interactions, where, for finite $L$, integrability-breaking terms must exceed a finite threshold to generate spectral correlations. 

The emergence of chaotic spectral statistics is related to the appearance of chaotic eigenstates, which, in turn, affect the structure of the matrix elements of few-body observables written in the energy eigenbasis. As the system size increases, chaotic eigenstates acquire increasingly random-vector-like properties with respect to few-body observables. Consequently, the diagonal elements $O_{nn}=\langle n|\hat{O}|n\rangle$ become smooth functions of energy, with decreasing eigenstate-to-eigenstate fluctuations, while the off-diagonal elements $O_{mn}=\langle m|\hat{O}|n\rangle$ ($m\neq n$) behave as random variables with approximately Gaussian distributions centered at zero. In Fig.~\ref{fig:FIG2}(d), we show the distribution of the off-diagonal matrix elements of the total magnetization in the $z$-direction, $\hat{M}^z = (1/L) \sum_i^L \hat{\sigma}^z_i$, obtained from eigenstates in the center of one of the most populated energy bands of $\hat{H}^{(\alpha)}$. The resulting distribution is well described by a Gaussian, reflecting the chaotic structure of the eigenstates.

\section{Validity of the band-resolved description}
\label{sec:Leakage}

In this section, we quantify the degree to which the exact eigenstates of $\hat{H}^{(\alpha)}$ for $\alpha\neq0$ remain confined to a single energy band. We show that, over a broad parameter regime, interband leakage remains perturbatively small. This provides a quantitative justification for describing the long-time properties of the system within a single energy band and for constructing band-resolved microcanonical ensembles.

To place this band-resolved description in the context of ETH, we first briefly review thermalization in isolated quantum systems. We introduce concepts and notations that are then used for the analysis of interband leakage in Sec.~\ref{subSecLeak} and for the discussion of fragmented eigenstate thermalization in the following section.

\subsection{Eigenstate Thermalization}
In the study of thermalization in isolated quantum systems with many interacting particles, the focus is on the long-time behavior of local few-body observables $\hat{O}$, following unitary evolution from an initial state, which we take here to be a pure state $|\Psi(0)\rangle$. A common approach is to consider a quantum quench, where $|\Psi(0)\rangle$ is an eigenstate of an initial Hamiltonian $\hat H_{\rm ini}$, whose parameters are then abruptly changed, so that the subsequent dynamics is governed by a final Hamiltonian $\hat H_{\rm f}$.

Expanding the initial state in the eigenbasis of the final Hamiltonian, $\hat H_{\rm f}|n\rangle=E_n|n\rangle$, as
\begin{align}
|\Psi(0)\rangle=\sum_n c_n |n\rangle,
\qquad
c_n=\langle n|\Psi(0)\rangle,
\end{align}
the expectation value of $\hat O$ evolves as
\begin{align}
\langle \hat O(t)\rangle
=
\sum_{n,m} c_n^*c_m
e^{i(E_n-E_m)t}
\langle n|\hat O|m\rangle.
\end{align}
For a nondegenerate energy spectrum, the off-diagonal contributions vanish upon infinite-time averaging, leading to
\begin{align}
    \langle \hat{O} \rangle^\mathrm{DE} = \sum_n |c_n|^2 \langle n|\hat{O}|n \rangle. 
\end{align}
Thus, when the system equilibrates, $\langle\hat O(t)\rangle$ remains close to this value for most times. The infinite-time average can also be expressed as an expectation value over the diagonal ensemble,
\begin{align}
\langle\hat O\rangle^{\rm DE} = {\rm Tr} \!\left[ \hat\rho^{\rm DE}\hat O \right],
\label{Eq:DE}
\end{align}
where
\begin{align}
\hat\rho^{\rm DE} = \sum_n |c_n|^2
|n\rangle\langle n| ,
\label{Eq:rhoDE}
\end{align}
is the diagonal ensemble density matrix.

%%%%%%%%%%%%%%%%%%%%%%%%%%%%%%%
\begin{figure*}
    \begin{center}
    \includegraphics[width = 1.8\columnwidth]{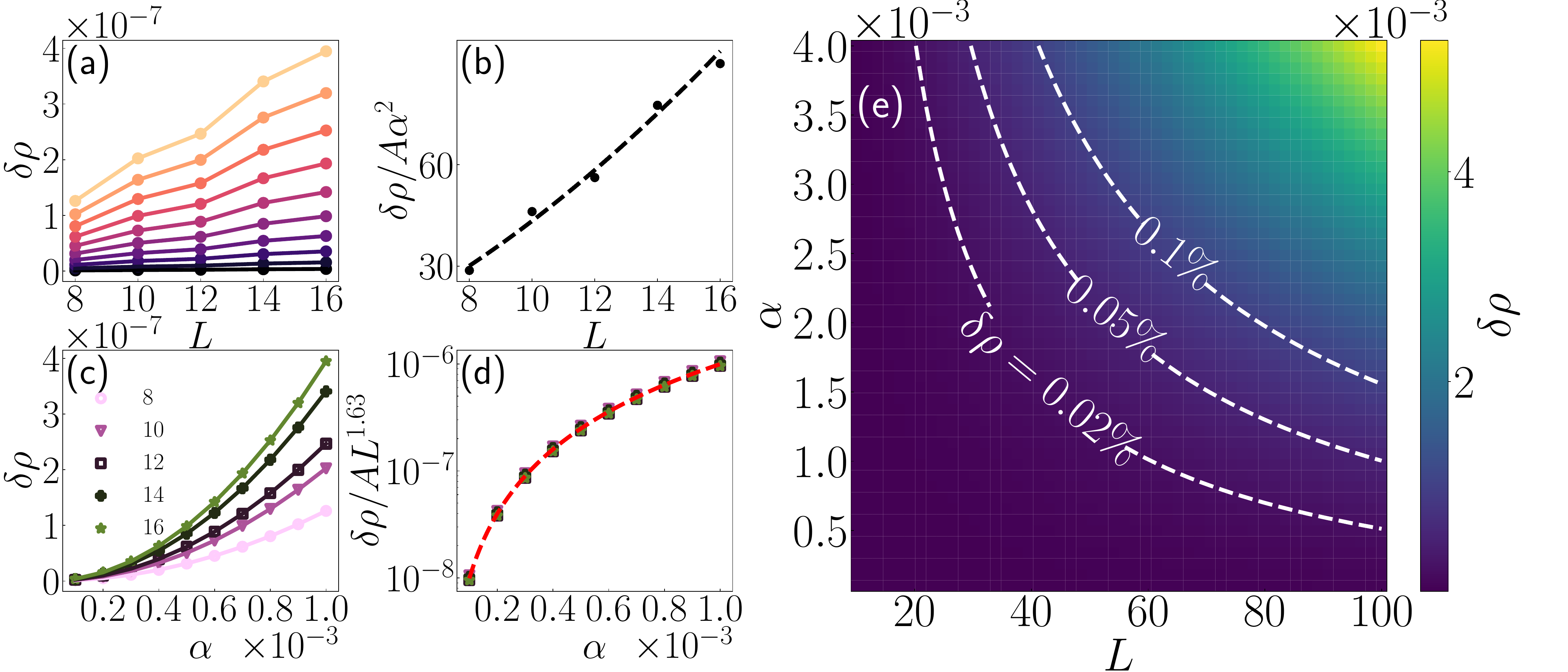}
        \caption{(a) Average leakage probability $\delta\rho$ as a function of system size $L$ for $\alpha$ ranging from $10^{-4}$ to $10^{-3}$ (darker to lighter colors). (b) Scaling collapse $\delta \rho \propto L^{1.57}$. (c) Average leakage $\delta \rho$ as a function of $\alpha$ for different system sizes as indicated in the legend. (d) Scaling collapse $\delta \rho(\alpha) \propto \alpha^{2.001}$.  The leakage is averaged over all eigenstates of $\hat H^{(\alpha)}$. (e) Density plot of leakage in the $(\alpha,L)$ plane, illustrating the region where the band-resolved description remains perturbatively accurate.
        }
        \label{fig:FIG5}
    \end{center}
\end{figure*}
%%%%%%%%%%%%%%%%%%%%%%%%%%%%%%%

As explained in Sec.~\ref{Sec:QChaos}, the onset of chaotic eigenstates implies that the diagonal matrix elements
$O_{nn}=\langle n|\hat O|n\rangle$ of few-body observables become smooth functions of energy and their eigenstate-to-eigenstate fluctuations decrease with increasing system size. As a result, the diagonal-ensemble value $\langle \hat{O} \rangle^\mathrm{DE}$ approaches the  microcanonical average,
\begin{align}
   O^{\rm mic}({\cal E}_0)
= \frac{1}{\mathcal N_{\Delta E}} \sum_{|E_n-{\cal E}_0| <\Delta E/2} \!\!\!\!  \!\!\!\! \langle n|\hat{O}|n \rangle ,
\end{align}
as the system size increases. In the equation above, $\Delta E$ denotes a suitably chosen energy window centered at the energy of the initial state,
\begin{align}
    {\cal E}_0 = \langle \Psi(0)|\hat{H}|\Psi(0)\rangle = \sum_n |c_n|^2 E_n ,
\end{align}
and $\mathcal{N}_{\Delta E}$ is the number of eigenstates inside this window. The convergence of $\langle \hat{O} \rangle^{\rm DE}$ toward $O^{\rm mic}({\cal E}_0)$, enabled by the emergence of chaotic eigenstates, is a central notion underlying ETH.

For generic many-body Hamiltonians with few-body interactions and interaction range exponent $\alpha>1$, the DOS typically has an approximately Gaussian profile encompassing the full spectrum. Chaotic eigenstates exist away from the edges of the spectrum. For this reason, studies of ETH commonly consider initial states with energies ${\cal E}_0$ within the bulk of the spectrum, often close to its center.

Unlike conventional ETH, the statistical description developed in this work is constructed within individual energy bands.
Although the band structure inherited from the fully connected limit persists for strong long-range interactions, the fragmentation is no longer exact. For $0<\alpha<1$, the conserved total-spin quantum number of the $\alpha=0$ Hamiltonian becomes only approximately conserved, allowing weak hybridization between different energy bands. As a result, the exact eigenstates acquire small perturbative admixtures from neighboring bands. Quantifying this interband leakage is essential for determining the regime in which the equilibrium properties remain accurately described by a single energy band and, consequently, by the fETH framework.

\subsection{Justification for band-resolved infinite-time averages and microcanonical ensemble}
\label{subSecLeak}

To quantify the confinement of the exact eigenstates to a single energy band, we compare an eigenstate $|n^{(0)}\rangle$, belonging to a given degenerate energy band of the parent Hamiltonian $\hat H^{(0)}$, with the eigenstates $|n\rangle$ of the perturbed Hamiltonian $\hat H^{(\alpha)}$, whose bands acquire a finite width. We define the leakage probability as the total weight lying outside the band $B_i$ of $\hat H^{(\alpha)}$ associated with the original band of the chosen state $|n^{(0)}\rangle$,
\begin{align}
\delta\rho(\alpha,L) = 1- \sum_{m\in B_i} |\langle m|n^{(0)}\rangle|^2.
\label{Eq:leakage}
\end{align}

Figures~\ref{fig:FIG5}(a) and \ref{fig:FIG5}(c) show that the leakage increases with both the exponent $\alpha$ and the system size $L$. Nevertheless, throughout the entire parameter regime explored numerically, $\delta \rho$ remains very small. Therefore, although the perturbation from $\hat{H}^{(0)}$ to $\hat{H}^{(\alpha)}$ couples different bands, the total probability transferred outside the original band remains negligible. This means that the equilibrium properties continue to be governed predominantly by the intraband structure, while interband hybridization contributes only perturbatively small corrections.

For the system sizes and interaction-range exponents considered here, the leakage is accurately described by
\begin{align}
    \delta\rho(\alpha,L)=A\,\alpha^{k_1}L^{k_2}
    \label{eq:Delta_rho}
\end{align}
with 
\[
A=4.41\times10^{-3},
\qquad
k_1=2.001,
\qquad
k_2=1.637.
\]
The scaling collapses shown in Figs.~\ref{fig:FIG5}(b) and \ref{fig:FIG5}(d) demonstrate that Eq.~\eqref{eq:Delta_rho} provides an excellent description of the numerical results over the fitted range. We therefore use Eq.~\eqref{eq:Delta_rho} as a practical diagnostic for estimating the region of the $(\alpha,L)$ plane where the band-resolved description remains quantitatively accurate. This information is summarized in Fig.~\ref{fig:FIG5}(e), which displays contours of constant leakage.

The perturbatively small interband leakage over a broad range of system sizes and  exponents $\alpha$ in the strong long-range regime [Fig.~\ref{fig:FIG5}(e)] implies that the observables equilibrium expectation values remain confined to individual energy bands, as we explain next.

Following a quench from $H_{\rm ini}=\hat{H}^{(0)}$ to $H_{\rm f}=\hat{H}^{(\alpha)}$, the infinite-time average of an observable, given by Eq.~(\ref{Eq:DE}), is determined by the diagonal ensemble in Eq.~(\ref{Eq:rhoDE}), where  $c_n=
\langle n|n^{(0)}\rangle $.
To quantify the contribution coming from a single energy band $B_i$, we define
the unnormalized band-resolved diagonal ensemble
\begin{align}
\hat\rho_{B_i}^{\rm DE} = \sum_{n\in B_i} |\langle n|n^{(0)}\rangle|^2 |n\rangle\langle n|.
\end{align}
Its trace is ${\rm Tr} \!\left[ \hat\rho_{B_i}^{\rm DE} \right] = 1-\delta\rho$, so that
\begin{align}
\left\| \hat\rho^{\rm DE} - \hat\rho_{B_i}^{\rm DE} \right\|_1 = \delta\rho.
\end{align}
Using the trace-norm inequality derived in Appendix~\ref{APP:Trace-Norm},
\begin{align}
\delta\langle\hat O\rangle^{\rm DE}&=\left| \langle\hat O\rangle^{\rm DE} - \langle\hat O\rangle_{B_i}^{\rm DE} \right|
\\
&= \left| {\rm Tr} \!\left[ \left( \hat\rho^{\rm DE} - \hat\rho_{B_i}^{\rm DE} \right) \hat O \right] \right|
\le \|\hat O\|_\infty \left\| \hat\rho^{\rm DE} -
\hat\rho_{B_i}^{\rm DE} \right\|_1, \nonumber
\end{align}
we obtain
\begin{align}
\delta\langle\hat O\rangle^{\rm DE} \le \|\hat O\|_\infty\, \delta\rho.
\label{eq:Delta_Obs}
\end{align}

Equation~\eqref{eq:Delta_Obs} shows that whenever $\delta\rho\ll1$, the exact diagonal-ensemble expectation value of a bounded local observable differs from its band-resolved value by at most a perturbatively small correction. Therefore, despite the weak hybridization between neighboring bands, equilibration remains effectively confined to a single energy band. 

This result has consequences for the statistical description of equilibrium. Since Sec.~\ref{sec:Model} established that the eigenstates are chaotic within each band, and the present analysis shows that the dynamics equilibrates within a single band, the equilibrium state must likewise be described by a statistical ensemble defined inside that same band. This provides the microscopic motivation for introducing a band-resolved microcanonical ensemble rather than a microcanonical ensemble constructed over the full spectrum.

The results obtained so far establish the physical foundation of fETH: exact eigenstates remain effectively localized within individual energy bands, equilibration occurs within those bands, and equilibrium is therefore governed by band-resolved statistical mechanics. The next section puts this picture to a quantitative test through the finite-size scaling of eigenstate expectation values inside
individual energy bands.

\section{Fragmented Eigenstate Thermalization}
\label{sec:MC}

Despite requiring exceedingly long timescales~\cite{Sriram2026_2}, the dynamics under $\hat{H}^{(\alpha)}$ with $0<\alpha \ll1$ eventually reaches equilibrium. We now investigate whether this equilibrium admits a statistical description.

For the present system with super-long-range interactions, eigenstates belonging to different energy bands cannot be treated on an equal footing in the construction of the microcanonical ensemble.  As discussed and justified in the previous section, the microcanonical ensemble is now defined within a single energy band,
\begin{align}
O^{\rm mic}_{B_i}({\cal E}_0)
= \frac{1}{\mathcal N_{B_i}} \!\!\!\! \!\!\!\! \sum_{\substack{n\in B_i\\ |E_n-{\cal E}_0| <\Delta E/2}} \!\!\!\!  \!\!\!\! O_{nn},
\label{eq:Band_MC}
\end{align}
where $B_i$ denotes the energy band dynamically connected to the initial state of energy ${\cal E}_0$, the microcanonical window is $\Delta E$, and $\mathcal N_{B_i}$ is the number of eigenstates in that window. 

Therefore, to establish that the infinite-time average of a few-body observable approaches the band-resolved microcanonical prediction $O^{\rm mic}_{B_i}({\cal E}_0)$ as the system size increases, we examine whether the eigenstate expectation values of the observable, $O_{nn}$, become smoother functions of energy within each band  as $L$ grows. Demonstrating this behavior validates the fETH. In the present system, however, this analysis first requires the proper identification of comparable energy bands across different system sizes, as discussed next.

\subsection{Selection of energy bands across system sizes}
\label{subsec:ZZtX}

To identify the broadened energy bands of $\hat H^{(\alpha)}$, we exploit the fact that they emerge  from the degenerate bands of the parent Hamiltonian $\hat H^{(0)}$. At $\alpha=0$, the total spin $s$ is conserved, and each $s$ sector gives rise to $(2s+1)$ energy bands. For fixed $s\ll L/2$, the energies of these parent bands approach the reference values
\begin{align}
E_m\simeq2hm-2J,
\qquad
m=-s,\ldots,s,
\label{Eq:Em}
\end{align}
because the interaction term contributes only corrections of order $s^2/L$. Here, $m$ should be understood as an asymptotic label for the parent bands in the fixed-$s$ limit, not as an exact quantum number of the finite-$L$ Hamiltonian.

Since the interband leakage is perturbatively small for the parameters considered (Sec.~\ref{sec:Leakage}), each broadened band of $\hat H^{(\alpha)}$ remains separated from neighboring bands and can be associated with a parent band of $\hat H^{(0)}$.  In practice, this identification is guided by the reference energies in Eq.~\eqref{Eq:Em}. A more precise way to identify the bands, based on projectors onto the parent-band subspaces, is described in Appendix~\ref{PreciseWay}. In Appendix~\ref{subsec:XXZ}, we also present another spin model, for which the reference energies can be obtained exactly as functions of the spin and magnetization quantum numbers $s$ and $m$, respectively.

The band identification must also respect the exact symmetries of the Hamiltonian. Parity partitions the spectrum into symmetric and antisymmetric sectors, while spin inversion imposes a selection rule that determines which parent bands can be compared across different system sizes, as explained below.

\begin{figure}[h]
    \begin{center}
        \includegraphics[width=1.0\columnwidth]{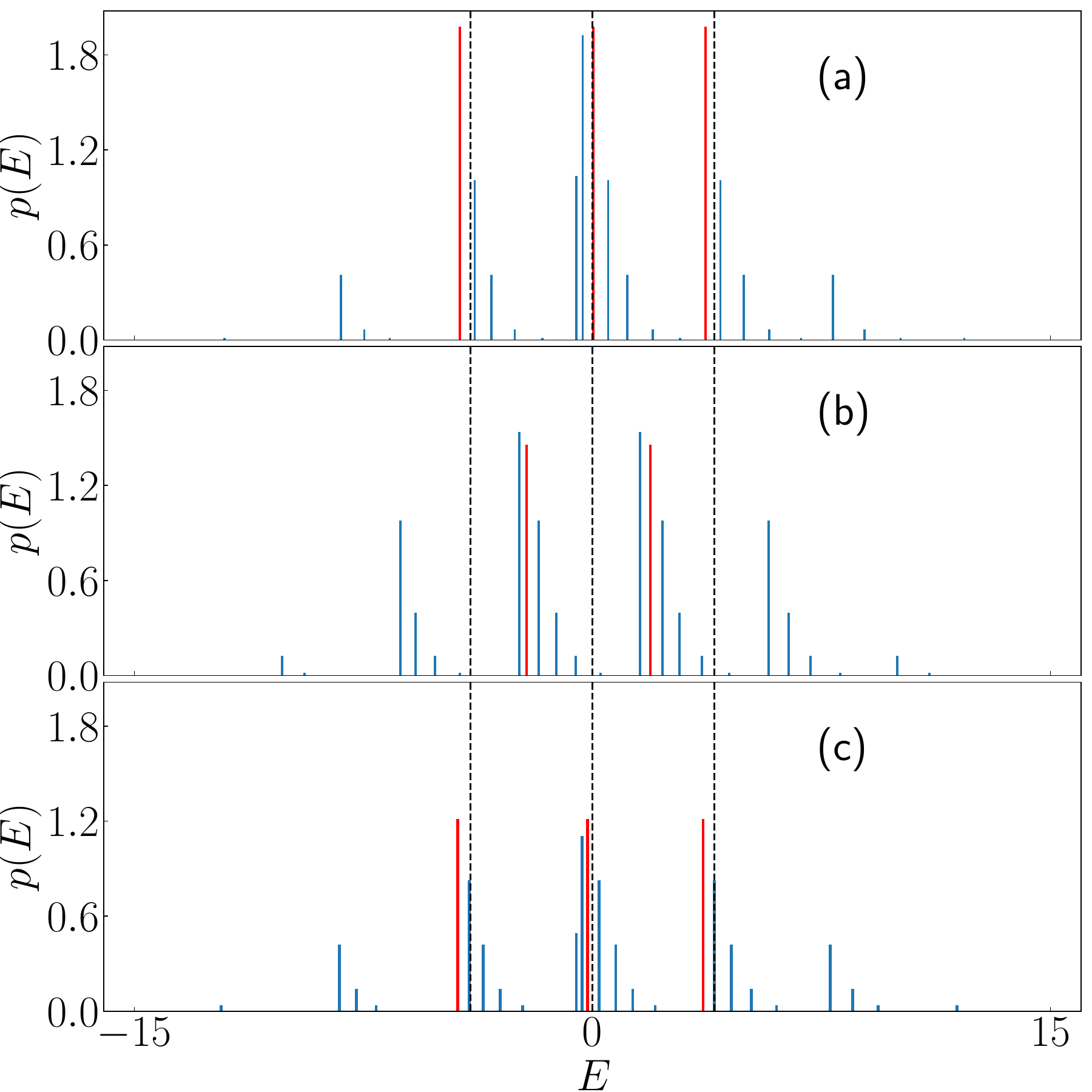}
        \caption{Density of states for $\alpha = 0$, spin inversion sector with $r = +1$, and system sizes (a) $L=12$, (b) $L=14$, and (c) $L=16$. The energy bands highlighted in red are those identified with  $s=2$. The vertical dashed black lines indicate the reference energies in Eq.~(\ref{Eq:Em}).  Due to spin-inversion symmetry, not all five bands associated with $s=2$ are present. The pattern of visible and depleted bands depends on the system size: the same set of bands is observed for $L=12$ and $L=16$, while a different subset appears for $L=14$, consistent with the $L \to L+4k$ selection rule. 
        }
        \label{fig:DOS_N121416}
    \end{center}
\end{figure}

In Fig.~\ref{fig:DOS_N121416}, we show the density of states for $\alpha = 0$ and system sizes (a) $L=12$, (b) $L=14$ and (c) $L=16$ for the spin inversion sector with $r=+1$. The bands highlighted in red are those associated with the $s=2$ sector. The vertical dashed black lines indicate the reference energies in Eq.~\eqref{Eq:Em} for $s=2$ and the even asymptotic labels $m=-2,0,2$.

For $s=2$, the parent Hamiltonian contains $(2s+1)=5$ energy bands. However, within a fixed spin-inversion sector, only a subset of these bands appears for a given system size. This is clearly illustrated in Fig.~\ref{fig:DOS_N121416}. For $L=12$ and $L=16$, only the bands associated with $m=-2,0,2$ are visible, whereas for $L=14$ we see the complementary pair, associated with $m=-1,1$. This difference follows from the spin-inversion selection rule, which depends on both $m$ and $L$.

As proven in Appendix~\ref{app:SelectionRule}, for even $L$,  the spin-inversion operator $\hat R^z=\prod_i\hat\sigma_i^z$ acts on the parent bands as  
\begin{align}
    \hat{R}^z |n^{(0)}\rangle = (-1)^{L/2 + m} |n^{(0)}\rangle .
\end{align}
Thus, for a fixed value of $m$ and a chosen inversion symmetry sector ($r=\pm 1$), changing the system size by $L\to L+2$ flips the spin-inversion eigenvalue, whereas changing it by $L\to L+4$ preserves it:
\begin{align}
    (-1)^{(L+2)/2 + m} &= -(-1)^{L/2 + m}, \\
    (-1)^{(L+4)/2 + m} &= (-1)^{L/2 + m}.
\end{align}
As a result, a parent band belongs to the same spin-inversion sector only for system sizes related by
\begin{align}
L\to L+4k, \qquad
k\in\mathbb Z.
\end{align}
This means that a consistent finite-size comparison of eigenstate expectation values for $\alpha \neq 0$  must be performed across such symmetry-compatible system sizes.

\begin{figure}
    \begin{center}
        \includegraphics[width=1.0\columnwidth]{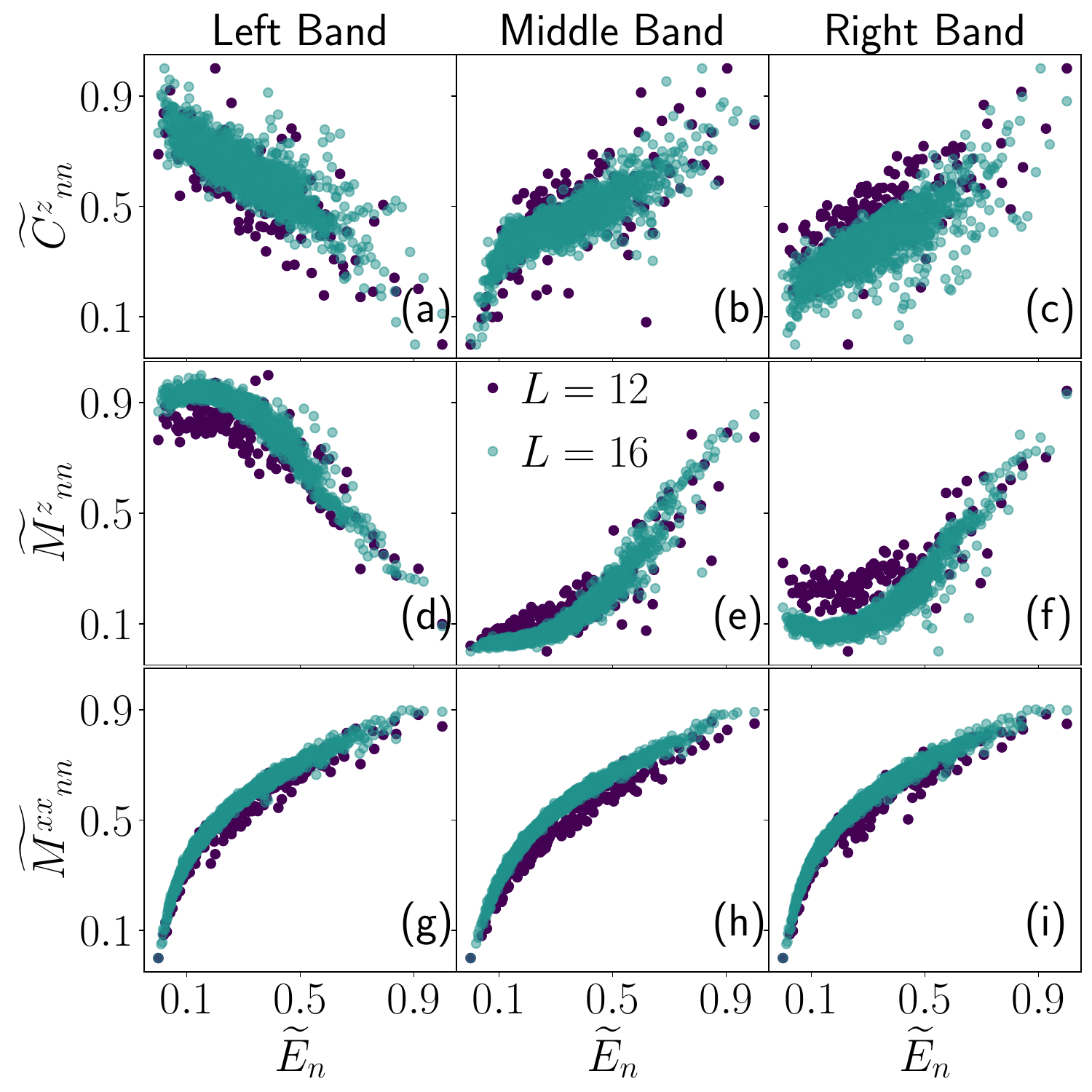}
        \caption{Rescaled eigenstate expectation values as a function of the rescaled energies for (a)-(c) first moment
of the excitation density $\hat{C}^z$, (d)-(f) total magnetization $\hat{M}^z$, and (g)-(i) spin-spin correlation function $\hat{M}^{xx}$. System sizes: $L=12$ (purple) and $L=16$ (teal); $\alpha = 10^{-4}$. Each column corresponds to one of the bands inherited from the three $s=2$ parent bands identified according to the procedure illustrated in Fig.~\ref{fig:DOS_N121416}. As $L$ increases, the fluctuations decrease within each band, consistent with fETH. 
}
        \label{fig:Micro_N1216}
    \end{center}
\end{figure}

\subsection{Eigenstate expectation values within fETH}

After identifying the appropriate bands, we rescale energies and eigenstate expectation values to the interval $[-1,1]$ to remove the differences in scale between different bands and system sizes. We work with
\begin{align}
    \widetilde{E}_n &= \frac{E_n - \min\{E_n\}}{\max\{E_n\} - \min\{E_n\}},\\
    \widetilde{O}_{nn} &= \frac{ O_{nn} - \min\{ O_{nn} \}}{\max\{O_{nn}\} - \min\{O_{nn}\}}.
    \label{eq:Renorm}
\end{align}

The rescaled eigenstate expectation values are shown in Fig.~\ref{fig:Micro_N1216} for the following three local observables:

\begin{itemize}

\item The observable introduced in the experiment on prethermalization in long-range
interacting spin chains~\cite{Neyenhuise2017}, which corresponds to the first moment of the excitation density,
\begin{align}
    C^z(t)
    &= \sum_{i=1}^{L}
    \langle\Psi(t)|x_i\hat{n}_i|\Psi(t)\rangle,
    \label{Eq:Cz}
\end{align}
where
\[
x_i=\frac{2i-L-1}{L-1}
\]
is the normalized position of lattice site $i$ and
\[
\hat{n}_i=(\hat{\sigma}_i^z+\hat{\mathbb I})/2
\]
is the local number operator.

\item The total magnetization along the $z$-axis,
\begin{align}
    M^z(t) &= \frac{1}{L} \sum_{i=1}^{L}
    \langle \Psi(t)| \hat{\sigma}_i^z | \Psi(t) \rangle.
    \label{Eq:Mz}
\end{align}

\item The nearest-neighbor spin--spin correlation function along the $x$-direction,
\begin{align}
    M^{xx}(t)
    = \sum_{i=1}^{L-1}
    \langle \Psi(t)|
    \hat{\sigma}_i^x \hat{\sigma}_{i+1}^x
    |\Psi(t)\rangle.
    \label{Eq:Mxx}
\end{align} 

\end{itemize}
Two system sizes are considered in the super-long-range regime.  The three columns correspond to the energy bands associated with the $s=2$ parent bands from the $\alpha=0$ limit. After accounting for parity and spin-inversion symmetry, as in Fig.~\ref{fig:DOS_N121416}, only three bands survive and the study is restricted to sizes differing by multiples of four, specifically $L=12$ (purple) and $L=16$ (teal).

As the system size increases from $L=12$ to $L=16$, the eigenstate expectation values become progressively smoother functions of energy, and their fluctuations decrease within each band. This behavior is observed consistently across all three bands and for all three observables. 

Combined with the leakage analysis of Sec.~\ref{sec:Leakage}, these results indicate that, throughout the parameter regime where interband leakage remains perturbatively small, the infinite-time averages of local observables approach the predictions of the corresponding band-resolved microcanonical ensemble. These findings provide numerical evidence in support of the fETH.

\begin{figure*}
    \begin{center}    
    \includegraphics[width=1.0\textwidth]{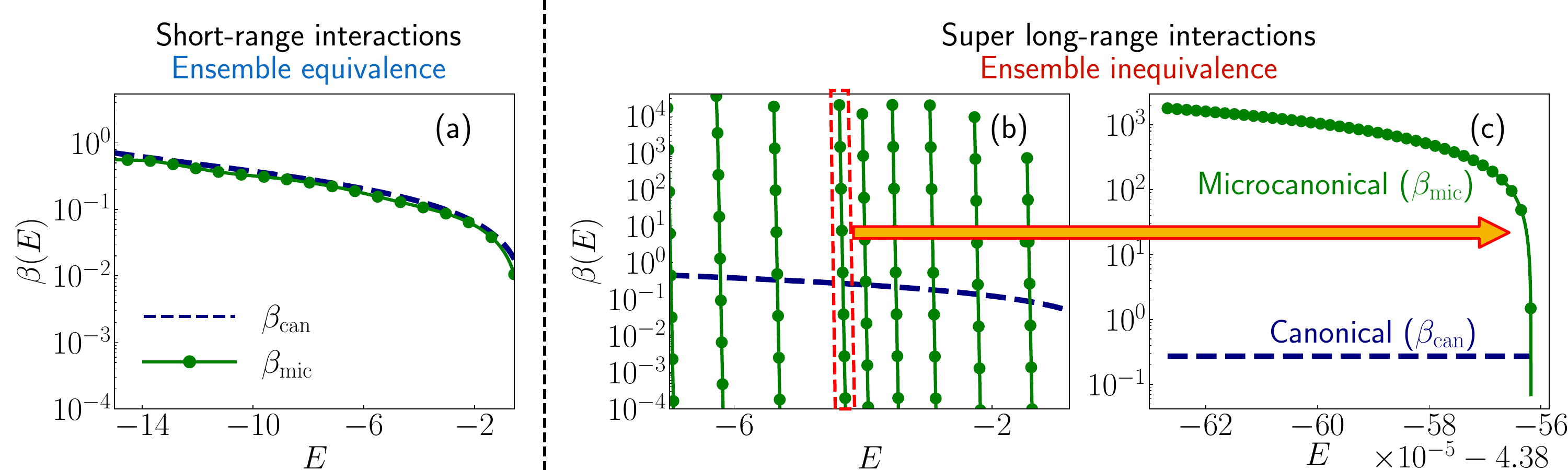}
        \caption{Inverse temperature for the canonical ($\beta_{\text{can}}$) and band-resolved microcanonical ($\beta_{\text{mic}}$) ensembles as a function of energy for $\hat{H}^{(\alpha)}$ with $L=16$. (a) Short-range interactions ($\alpha = 3$): away from the spectral edges, $\beta_{\mathrm{can}}$ and $\beta_{\mathrm{mic}}$ coincide, indicating ensemble equivalence.
        (b)-(c) Super long-range interactions ($\alpha = 10^{-4}$): ensemble inequivalence. (b) The canonical inverse temperature varies smoothly with energy, whereas the same values of $\beta_{\rm mic}$ reappear at different energies because the spectrum is fragmented into energy bands. (c) Enlargement of one energy band. While $\beta_{\rm can}$ is nearly constant over the band, the band-resolved microcanonical inverse temperature $\beta_{\mathrm{mc}}$ varies substantially across it. 
        }
        \label{fig:FIG9}
    \end{center}
\end{figure*}

\section{Ensemble inequivalence}
\label{sec:Ensemble_Inequivalence}

We now explore the equilibrium consequences of the band-resolved statistical description. In particular, we compare the band-resolved microcanonical ensemble with the canonical ensemble. We show that, throughout the parameter regime where interband leakage remains perturbatively small, the two ensembles yield different predictions.

Unlike conventional approaches, where ensemble inequivalence is associated with nonconcave entropy and equilibrium phase transitions, we identify a microscopic origin for it. It arises from the band structure of the spectrum. As demonstrated in the previous two sections, the appropriate microcanonical ensemble must be constructed within individual energy bands, whereas the canonical ensemble weighs eigenstates solely according to their energies and therefore mixes contributions from different bands. Since the two ensembles probe different regions of Hilbert space, they generally lead to different equilibrium predictions.

Ensemble inequivalence was first discussed in classical statistical mechanics, where the equivalence between microcanonical (fixed energy) and canonical (fixed temperature) ensembles relies on additivity.  In additive systems, the canonical ensemble effectively samples a narrow energy shell around the microcanonical energy, so the two ensembles become equivalent in the thermodynamic limit. For systems with long-range interactions, however, additivity breaks down. As a consequence, the canonical ensemble may mix states that are not accessible within a single microcanonical shell, leading to inequivalent predictions. This
phenomenon is prominent near first-order phase transitions and is commonly associated with nonconcave entropy, multivalued caloric curves, and negative specific heat,~\cite{Lynden-Bell1967,Hertel1971,Thirring2003,Barre2001,Campa2009,Campa2014}.

In quantum systems, ensemble inequivalence has been investigated in the context of long-range interacting models and quantum phase transitions. The standard approach compares canonical and microcanonical phase diagrams~\cite{Defenu2024_1,ArrufatVicente2025}, so the mechanism is still thermodynamic. 

\subsection{Band-resolved microcanonical ensemble versus canonical ensemble}

We now show that fETH provides a different microscopic route to ensemble inequivalence that already appears in finite systems.

To compare the two ensembles, we extract the inverse temperature from each of them.  Since the microcanonical ensemble is now defined within an individual energy band, its inverse temperature is defined as 
\begin{equation}
\beta_{\mathrm{mic}}(E) = \frac{\partial S_{B_i}(E)}{\partial E},
\end{equation}
where $S_{B_i}(E)$ denotes the entropy associated with the density of states inside the selected energy band $B_i$. For the canonical ensemble, the inverse temperature is fixed by construction and the corresponding average energy is
\begin{equation}
E(\beta_{\mathrm{can}}) = \frac{\mathrm{Tr}\left[ \hat{H} e^{-\beta_{\mathrm{can}} \hat{H}} \right]}{\mathrm{Tr}\left[ e^{-\beta_{\mathrm{can}} \hat{H}} \right]},
\end{equation}
which can be inverted to obtain $\beta_{\mathrm{can}}(E)$.

The comparison between $\beta_{\mathrm{mic}}(E)$ and $\beta_{\mathrm{can}}(E)$ is shown in Fig.~\ref{fig:FIG9}. For short-range interactions [Fig.~\ref{fig:FIG9}(a)], where $\alpha=3$, the spectrum is not fragmented and the two temperatures coincide, indicating ensemble equivalence. Small deviations appear only near the spectral edges, where finite-size effects are enhanced. The figure displays only negative energies up to $E=0$, which corresponds to the infinite-temperature limit.

The situation changes qualitatively for strong long-range interactions, as shown in Fig.~\ref{fig:FIG9}(b). In this regime, the spectrum fragments into distinct energy bands. The canonical ensemble weighs eigenstates through the Boltzmann factor $e^{-\beta_{\rm can}E_n}$ and therefore averages over all energy bands without distinguishing their fragmented origin. Therefore, the canonical inverse temperature remains a smooth function of energy. By contrast, the microcanonical temperature must be extracted separately within each band. Since the bands have similar internal density of states, the same values of $\beta_{\rm mic}$ reappear at different energies, producing the repeated vertical structures seen in Fig.~\ref{fig:FIG9}(b).

Figure~\ref{fig:FIG9}(c) illustrates the behavior within a single energy band. Over this narrow energy window, $\beta_{\rm can}$ is almost constant because the canonical ensemble is insensitive to the internal structure of an isolated band. The band-resolved microcanonical inverse temperature, however, varies significantly across the same interval, reflecting the skewed-Gaussian density of states inside the band [Fig.~\ref{fig:FIG1} (b)]. 

The resulting ensemble inequivalence is therefore fundamentally different from the conventional thermodynamic mechanism. The canonical and microcanonical ensembles assign different temperatures to the same set of states. The resulting inequivalence arises  from the fragmented structure of the spectrum for a finite system size and the need to define the microcanonical ensemble within individual bands.

\section{Conclusions}
\label{sec:Conclusion}

We have developed a theoretical framework for understanding thermalization and ensemble inequivalence in the broad class of quantum many-body systems with approximate full permutation symmetry. Our results are demonstrated through mathematical analysis and extensive numerical simulations of a finite spin-$1/2$ model with strong long-range interactions.

We have shown that thermalization can persist in these systems despite the fragmentation of the many-body spectrum into weakly coupled energy bands. In the fully connected limit, full permutation symmetry gives rise to highly degenerate bands. As soon as the system departs from this limit, however, the degeneracies are lifted and quantum chaos develops within individual bands, while hybridization between different bands remains perturbatively weak over a broad parameter regime. We quantified this interband leakage and established its scaling with the interaction-range exponent and system size, thereby identifying the regime in which equilibration remains effectively confined to individual energy bands.

The separation between strong intraband chaos and weak interband hybridization provides the basis for a band-resolved description of thermalization, referred to as fragmented eigenstate thermalization (fETH). Rather than constructing a microcanonical ensemble from the full spectrum, fETH associates equilibrium with the energy band dynamically connected to the initial state. %Within this band, eigenstate expectation values of local observables become progressively smoother with increasing system size.

Testing fETH requires abandoning the conventional finite-size scaling used in short-range systems. We proved a symmetry-imposed selection rule that determines the sequences of system sizes that can be meaningfully compared. Permutation symmetry, together with spin-inversion symmetry, restricts comparisons to sequences of system sizes $L,L+4,L+8,\ldots$. Along these sequences, eigenstate-to-eigenstate fluctuations decrease with increasing system size, supporting the validity of fETH.

The band-resolved structure of thermalization also has consequences for equilibrium statistical mechanics. The microcanonical ensemble associated with fETH samples states within a single energy band, whereas the canonical ensemble weighs states according to energy without resolving their band origin and therefore mixes contributions from different bands. The two ensembles consequently explore different regions of the Hilbert space and yield inequivalent temperature-energy relations. This reveals a spectral origin of ensemble inequivalence in finite systems, rooted in the fragmented structure of the many-body spectrum rather than in the equilibrium phase transitions or nonconcave entropy conventionally associated with ensemble inequivalence in long-range systems.

In summary, our results establish a framework for thermalization in finite many-body quantum systems close to a fully permutation-symmetric limit. In the regime where interband hybridization remains perturbatively weak, Hilbert-space fragmentation and thermalization are not mutually exclusive. Global ergodicity is lost, yet quantum chaos and thermalization emerge within individual energy bands. Approximate permutation symmetry provides the unifying mechanism connecting spectral fragmentation, band-resolved thermalization, and ensemble inequivalence.

\begin{acknowledgements}
C.L.S. and L.F.S are supported by UConn funds. S.K.P is supported in TIFR through a graduate fellowship from the Department of Atomic Energy  (D.A.E), Govt. of India. 
\end{acknowledgements}

%\section*{Data Availability}
%The data that support the findings of this study are openly available on Zenodo with the link in Ref.~\cite{DatafETH}.

%%%%%%%%%%
\appendix
%%%%%%%%%%

\section{Trace-norm inequality} \label{APP:Trace-Norm}
\textit{\textbf{Lemma}}:---
Let $A$ be a bounded operator and $X$ be a trace-class operator on a finite-dimensional Hilbert space. Then
\begin{equation}
|\mathrm{tr}(A X)| \le \|A\|_{\infty} ~ \|X\|_1,
\end{equation}
where $\|A\|_{\infty} = \max\{Spec(A)\}$ is the operator norm (maximum eigenvalue) and $\|X\|_1 = Tr[\sqrt{X_1^\dagger X_1}]$ is the trace norm.

\textit{\textbf{Proof}}:---
Let $X$ have the singular value decomposition (SVD)
\begin{equation}
X = U \Sigma V^\dagger,
\end{equation}
where $U$ and $V$ are unitary matrices, and $\Sigma = \mathrm{diag}(s_1, s_2, \dots, s_n)$ with $s_i \ge 0$ the singular values of $X$. Then
\begin{align}
|\mathrm{tr}(A X)| &= |\mathrm{tr}(A U \Sigma V^\dagger)| \\
&= |\mathrm{tr}(V^\dagger A U \, \Sigma)| \\
&= \left| \sum_i s_i (V^\dagger A U)_{ii} \right|.
\end{align}

Applying the triangle inequality and using the definition of the operator norm $\|W\|_{\infty} := \sup_{\|v\|=1} \|W v\|$, we get
\begin{equation}
\left| \sum_i s_i (V^\dagger A U)_{ii} \right| \le \sum_i s_i |(V^\dagger A U)_{ii}| \le \sum_i s_i \, \|V^\dagger A U\|_{\infty}.
\end{equation}

Since the operator norm is unitarily invariant,
\begin{equation}
\|V^\dagger A U\|_{\infty} = \|A\|_{\infty},
\end{equation}
we obtain
\begin{equation}
|\mathrm{tr}(A X)| \le \sum_i s_i \, \|A\|_{\infty} = \|A\|_{\infty} \sum_i s_i = \|A\|_{\infty} ~ \|X\|_1.
\end{equation}

This completes the proof.

%%%%%%%%%%%%%%%%%%%%%%%
\section{Projector-based identification of broadened energy bands}
\label{PreciseWay}

In the main text, broadened energy bands of $\hat H^{(\alpha)}$ are identified using their proximity to the reference energies of the parent Hamiltonian $\hat H^{(0)}$, together with the exact parity and spin-inversion symmetries. This procedure is sufficient in the parameter regime considered here, where interband leakage is small and the broadened bands remain clearly resolved. In this Appendix, we describe a more precise, basis-independent way to identify the bands.

At $\alpha=0$, the Hamiltonian $\hat H^{(0)}$ conserves the total spin $s$, and its spectrum is organized into degenerate energy bands within each $s$ sector. A given parent band is not associated with a unique eigenstate. Instead, it is a degenerate subspace. If $\mathcal B^{(0)}$ denotes one such parent band, we define the projector onto this subspace as
\begin{align}
\hat P_{\mathcal B}^{(0)}
=
\sum_{a\in\mathcal B^{(0)}}
|a^{(0)}\rangle\langle a^{(0)}|,
\label{eq:ParentProjector}
\end{align}
where the states $|a^{(0)}\rangle$ form any orthonormal basis of the degenerate parent-band subspace. The label $a$ simply enumerates the states inside the degenerate band. The projector in Eq.~\eqref{eq:ParentProjector} is independent of the particular basis chosen inside the degenerate subspace.

Since $\hat H^{(0)}$ also commutes with parity and spin inversion, the basis states $|a^{(0)}\rangle$ can be chosen to have definite parity and spin-inversion eigenvalues. Equivalently, one may define $\hat P_{\mathcal B}^{(0)}$ directly within a fixed symmetry sector. This is the most convenient choice when comparing with the eigenstates of $\hat H^{(\alpha)}$, which are also computed within fixed parity and spin-inversion sectors.

For an eigenstate $|n\rangle$ of the perturbed Hamiltonian $\hat H^{(\alpha)}$, we define its weight in the parent band $\mathcal B^{(0)}$ as
\begin{align}
w_n^{\mathcal B}
=
\langle n|
\hat P_{\mathcal B}^{(0)}
|n\rangle .
\label{eq:BandWeight}
\end{align}
This quantity measures how much of the exact eigenstate $|n\rangle$ lies inside the parent-band subspace. It is invariant under arbitrary rotations of the degenerate eigenbasis of $\hat H^{(0)}$ and is therefore the appropriate object to use when the parent band is degenerate.

The broadened band $\mathcal B$ of $\hat H^{(\alpha)}$ associated with the parent band $\mathcal B^{(0)}$ can then be defined as the set of eigenstates $|n\rangle$ for which $w_n^{\mathcal B}$ is maximal among all parent-band projectors in the same symmetry sector. In the perturbative regime where interband leakage is small,
\begin{align}
w_n^{\mathcal B}\simeq 1
\end{align}
for eigenstates belonging to the corresponding broadened band, while the weights in other parent bands are small. In this regime, the projector-based identification is unambiguous and coincides with the visible band clusters in the density of states.

This construction also clarifies why overlaps with individual eigenstates of $\hat H^{(0)}$ are not sufficient. Because the parent bands are degenerate, individual eigenstates inside a band are not uniquely defined: any orthonormal rotation within the degenerate subspace gives an equally valid eigenbasis. The projector $\hat P_{\mathcal B}^{(0)}$, by contrast, is basis independent and uniquely represents the parent band as a subspace.

The reference energies in Eq.~\eqref{Eq:Em} are therefore best understood as practical markers for locating the broadened bands in the density of states. The projector weights in Eq.~\eqref{eq:BandWeight} provide the corresponding basis-independent criterion for assigning exact eigenstates of $\hat H^{(\alpha)}$ to their parent bands.

\section{XXZ chain}
\label{subsec:XXZ}
The Hamiltonian we wish to consider is,
\begin{align}
    \hat{H}^{(0)}_\mathrm{XXZ}  = \frac{J}{N} \sum_{i = 1}^{N} \sum_{j >i} \biggl( \hat{\sigma}_i^x \hat{\sigma}_j^x +\hat{\sigma}_i^y \hat{\sigma}_j^y + \Delta \hat{\sigma}_i^z \hat{\sigma}_j^z \biggr)- h \sum_{i=1}^{N} \hat{\sigma}_i^z,
    \label{eq:H_XXZ}
\end{align}
where all variables have the same meaning as in Eq.~\eqref{eq:XXtZH}. The energy eigenvalues of the Hamiltonian may be written out explicitly as,
\begin{align}
    E(s, m) &= \frac{J}{2N}s(s+1) + \frac{2J}{N}(\Delta-1)m^2 \nonumber \\
    & \qquad \qquad- J (2+\Delta) - 2hm,
    \label{eq:E_XXZ}
\end{align}
where $(s,m)$ are eigenvalues of $\hat{S}$ and $(1/2) \sum_{i =1}^L \hat{\sigma}_i^z$, respectively. We immediately see that in the limit $N \to \infty$, the energy for a fixed $(s,m)$ becomes,
\begin{align}
    E_{(s,m)} = -J(2+\Delta) - 2hm,
    \label{eq:EInf_XXZ}
\end{align}
where there are $(2s+1)$ energy eigenvalues corresponding to the multiplicity of the $s$-sector. The most important aspect of Eq.~\eqref{eq:EInf_XXZ} is that for a given $(s,m)$, $E_{(s,m)}$ is independent of the system size $N$. Thus, the $(s,m)$ band asymptotically approaches the energy $E_{(s,m)}$, with increasing system size.

\section{Selection rule for finite-size scaling}
\label{app:SelectionRule}

\textbf{Lemma:-} Consider a Hamiltonian $\hat{H}$ on $L$ spin-$\tfrac{1}{2}$ degrees of freedom with the Hilbert space $(\mathbb{C}_2)^{\otimes L}$. If $[\hat{H}, \hat{S}^2] = [\hat{H}, \hat{R}^z] = 0$, where $\hat{S}^2$ is the total spin, and $\hat{R}^z$ ($(\hat{R}^z)^2 = \hat{\mathbb{I}}$) is the $\mathbb{Z}_2$ spin-inversion symmetry, then $\hat{H}$, $\hat{S}^2$, and $\hat{R}^z$ share a common eigenbasis. Consequently, any total spin sector $\mathcal{H}_s$ partitions into two invariant subspaces $\mathcal{H}_{s, \pm}$ defined by the eigenvalues $\mathcal{R} = \pm 1$. These are spanned by the basis $\{|s, m\rangle\}$ where $m$ is restricted to either purely even or purely odd values. Furthermore, for even $L$, the symmetry blocks $(s, \mathcal{R})$ exhibit a periodicity of $L \to L+4$.\\

\textbf{Proof:-} Let us first fix the choice of  basis to be $|s,m\rangle$ such that we have 
\begin{align}
    \hat{S}^2 |s,m\rangle = s(s+1) |s,m\rangle , ~\hat{S}_z |s,m\rangle= m|s,m\rangle, 
\end{align}
where $\hat{S}^z = (1/2)\sum_{i=1}^L \hat{\sigma}^z$. Next, we always fix the the inversion operation in the direction of Zeeman splitting, represented as $\hat{R}^z = \prod_{i=1}^L \hat{\sigma}_i^z$. Note, that this choice of direction is arbitrary, and is therefore invariant under any rotation of the Zeeman field.

We can express the exponential of $\hat{S}^z$ as,
\begin{align}
    e^{-i \pi \hat{S}^z} = \prod_{i=1}^L e^{-i \frac{\pi}{2} \hat{\sigma}_i^z} = (-i)^L \prod_{i=1}^{L} \hat{\sigma}_i^z.
    \label{eq:ExpSz}
\end{align}
Using Eq.~\eqref{eq:ExpSz}, we rewrite our spin-inversion operator as,
\begin{align}
    \hat{R}^z = (-1)^\frac{-L}{2} e^{-i \pi \hat{S}^z},
    \label{eq:Iz}
\end{align}
wherein we have omitted a factor of $(-1)^L$ because we will set our choice of $L$ to be even.

Since $[\hat{S}^z, \hat{S}^2]=0$, we have $[\hat{R}^z,\hat{S}^2]=0$, implying $\hat{H},~\hat{S}^2, ~\hat{R}^z$ share a common eigenbasis.

Now, the manifold $\mathcal{H}_s$ corresponding to a given total spin $s$ is spanned by the states $\{|s,m\rangle\}_{m=-s,\ldots,0,\ldots,s}$. Acting the inversion on these basis states and using Eq.~\eqref{eq:Iz}, we obtain
\begin{align}\label{eq:Inversion eigenvalue relation}
    \hat{R}^z |s,m\rangle = (i)^L e^{i\pi m} |s, m\rangle = (-1)^{\frac{L}{2} +m} |s,m\rangle \equiv \mathcal{R} |s,m \rangle, 
\end{align}
implying $\hat{R}^z$ is diagonal in the chosen basis. Moreover, this relation reveals that, for a choice of $L/2$, these basis states can be grouped into either all $m$ being even, or all $m$ odd with distinct inversion eigenvalue. This leads to a block decomposition of the manifold $\mathcal{H}_s$, for example, if we take $L/2$ to be even, then all the states $\{|s,m\rangle, \forall~m \in even\}$ will have $\mathcal{R}=+1$ and $\{|s,m\rangle, \forall~m \in odd\}$ will have $\mathcal{R}=-1$, leading to $\mathcal{H}_s= \underbrace{Span\left( \{|s,m\rangle, \forall~m \in even\} \right)}_{\mathcal{H}_{s,+}} \bigoplus \underbrace{Span\left( \{|s,m\rangle, \forall~m \in odd\} \right)}_{\mathcal{H}_{s,-}} $. Since $[\hat{H}, \hat{R}^z] = 0$, these subspaces are invariant under $\hat{H}$. Therefore, all eigenstates of $\hat{H}$ within a fixed spin sector $s$ lie entirely within one of the two inversion sectors.  

Finally, the symmetry sector $(s, \mathcal{R}) \equiv (s, (-1)^{L/2 + m})$ depends on the system size $L$. Requiring the symmetry sector to remain consistent with increasing $L$ necessitates:
\begin{align}
(-1)^{(L+4)/2 + m} = (-1)^{L/2 + 2 + m} = (-1)^{L/2 + m}.
\end{align}
Thus, the symmetry block $(s, (-1)^{L/2 + m})$ exhibits a periodicity of $L \to L+4$. (QED)

\section{Calculation of microcanonical temperature using kernel density estimation} \label{APP:KDE}
The microcanonical inverse temperature is defined as,
\begin{align}
    \beta_\mathrm{mic} = \frac{\mathrm{d} \log \Omega(E)}{\mathrm{d} E},
    \label{eq:Beta_mc}
\end{align}
where $\Omega(E)$ in the microcanonical picture is the number of microstates (here, eigenstates) with energy $\leq E$, which is obtained using the integral,
\begin{align}
    \Omega(E) = \int_{E_{\mathrm{min}}}^{E} g(\tau) ~\mathrm{d} \tau,
    \label{eq:DoS}
\end{align}
where $g$ is the density of states. Using Eq.~\eqref{eq:DoS} in Eq.~\eqref{eq:Beta_mc} we get,
\begin{align}
    \beta_\mathrm{mic} = \frac{\mathrm{d} \log g(E)}{\mathrm{d} E}.
    \label{eq:Beta_mc_2}
\end{align}
Thus the inverse temperature can be calculated from the density of states, provided it is a differentiable function of the energies. Detailed below is the method to obtain a differentiable estimate of the density of states from the eigenvalues of the Hamiltonian.

Consider Hamiltonian $\hat{H}$ with eigenvalues $\{ E_1, E_2, \dots, E_\mathcal{D} \}$ where $\mathcal{D}$ is the dimension of the Hilbert space. The histogram of these eigenvalues, represented by the function,
\begin{align}
    H(E) = \frac{1}{nb} \sum_{i=1}^\mathcal{D} 1_{|E-E_i| < b}
\end{align}
where $b$ is the bin width and $n$ is the number of bins, gives a discontinuous approximation of the density of states of $\hat{H}$ . A smoothened density of states may be obtained by a kernel density estimator (KDE)~\cite{Parzen1962KDE, Rosenblatt1956KDE, Schmidt2022KDE, Scott1992Book, Sui2014KDE} as,
\begin{align}
    \hat{g}(E) &= \frac{1}{\mathcal{D}h} \sum_{i=1}^{\mathcal{D}} K \biggl(\frac{E-E_i}{h} \biggr) \nonumber \\ 
    &= \frac{1}{\mathcal{D}} \sum_{i=1}^{\mathcal{D}} K_h (E- E_i),
\end{align}
with $K_h(x) = (1/h) K(x/h)$ and $h$ quantifies the width of the kernel function $K$, a popular choice for which is the Gaussian. Thus a KDE is a normalized pdf which is a collection of peaked Gaussian functions centered at the data points. In the limit $h \to 0, \mathcal{D}h \to \infty; \hat{g}(E) \to g(E)$, where $g(E)$ is the true density of states. To ensure normalization of $\hat{g}(E)$, we have,
\begin{align}
    \int K(\tau) ~\mathrm{d} \tau = 1; ~K(\tau) \geq 0.
\end{align}

To show the existence of the derivative, we show that the KDE is a convolution. The distribution of the data points may be written as the following sum of $\delta$ functions,
\begin{align}
    g_e(E) = \frac{1}{\mathcal{D}} \sum_{i=1}^{\mathcal{D}} \delta(E-E_i)
\end{align}

The KDE is,

\begin{align}
    \hat{g}(E) & = \frac{1}{\mathcal{D}} \sum_i K_h(E-E_i)\\
    & = \frac{1}{\mathcal{D}} \sum_{i=1}^{\mathcal{D}} \int \mathrm{d} \tau ~\delta (\tau - E_i) ~K_h (E-\tau) \\
    & = g_e(E) \circledast K_h(E).
\end{align}
This guarantees the existence of the derivative because the derivative of the convolution may be written as,
\begin{align}
    \frac{\mathrm{d} \hat{g}(E)}{\mathrm{d}E} = g_e(E) \circledast \frac{\mathrm{d}}{\mathrm{d}E} K_h (E).
\end{align}

We can now calculate the microcanonical inverse temperature as the logarithmic derivative of the density of states with respect to energy.

%%%%%%%%%%%%%%%%%%%%%%%%%%%%%%%%%%%%%%%%%%%%%%%%%%%%%%%%%%%%%
%%%%%%%%%%%%%%%%%%% REFERENCES %%%%%%%%%%%%%%%%%%%%%%%%%%%%%%
%%%%%%%%%%%%%%%%%%%%%%%%%%%%%%%%%%%%%%%%%%%%%%%%%%%%%%%%%%%%%
\bibliography{biblio2025}
%%%%%%%%%%%%%%%%%%%%%%%%%%%%%%%%%%%%%%%%%%%%%%%%%%%%%%%%%%%%%

\end{document}